\documentstyle[12pt]{article}

\begin{document}

{\bf Preprint SB/F/98-252} \vskip .5cm \hrule
\vskip 5.cm

\centerline{{\Large New Topological Aspects of $BF$ Theories}}

\vskip.5cm \centerline{{\large M. I. Caicedo and A. Restuccia}} \vskip .5cm %
\centerline{\it Universidad Sim\'{o}n Bol\'{\i}var, Departamento de
F\'{\i}sica} \vskip .5cm \centerline{\it Apartado postal 89000, Caracas
1080-A, Venezuela.} \centerline{\it e-mail: mcaicedo@fis.usb.ve,
arestu@usb.ve}

\vskip 1cm {\bf Abstract}

\begin{quotation}
\noindent {\small $BF$ theories defined over non trivial line bundles are
studied. It is shown that such theories describe a realization of a non
trivial higher order bundle. The partition function differs from the usual
one -in terms of the Ray Singer Torsion- by a factor that arises from the
non triviality of the line bundles.}
\end{quotation}

\vskip 1cm Topological field theories were introduced in \cite{re:Horowitz}
and \cite{re:BlauThompson1}\cite{re:BlauThom2}. The partition functions of
the abelian BF theories is a topological invariant of the base manifold $X$,
related to the Ray Singer torsion as shown by A. Schwarz in \cite{re:Schwarz}%
. Other interesting observables may be constructed as correlation functions
of Wilson surfaces associated to the $A$ and $B$ fields, these correlation
functions determine the linking and intersection numbers of manifolds of
several dimensions.

The BRST gauge fixing of the $BF$ action was studied for the abelian case in
\cite{re:BlauThom2} and completed for the non abelian case in \cite
{re:MarAlv1} and \cite{re:MarAlv2}, and several interesting properties of
the quantum effective action were analysed.

The interest on the $BF$ theories is not limited to the topological field
theory, it happens that the $BF$ lagrangian density appears as an
interaction term in several physical theories with local propagating degrees
of freedom \cite{re:PariasR} \cite{re:Stephany} in particular, as the
interaction term relating dual theories of $p$ and $(d-p-2)$-forms in $d$
dimensions \cite{re:MarAlvIsbe}, consequently the topological contribution
comming from $BF$ theories appear even in those physical theories with non
trivial physical hamiltonian.

In the analysis of \cite{re:Horowitz}\cite{re:BlauThompson1}\cite
{re:BlauThom2} the connection $1$-form $A$ of the $BF$ action is defined
over a flat vector bundle, in this note we will consider $BF$ theories over
non trivial $U(1)$ bundles. More precisely, we will assume the base $X$ to
be a compact, orientable finite dimensional euclidean manifold without
boundary, and will consider all the isomorphism classes of $U(1)$ line
bundles with connections that can be built over $X$. We will compare the
resulting theory with the standard one, that is, we will compare a $BF$
theory over a trivial line bundle with a $BF$ theory formulated over a non
trivial one. We will show that in the latest case the theory describes a
realization of a non trivial higher order bundle \cite{re:MarAlvIsbe}\cite
{re:Freed}\cite{re:Nepome}\cite{re:Brylinski}.

The partition function of both formulations have a common factor expressed
in terms of the Ray Singer torsion, the new theory however contains an extra
topological factor arising solely from the non triviality of the line
bundles themselves, which distingushes both theories, this latest factor is
introduced by the zero modes of the quantum effective action of the theory.
These non trivial contributions to the partition function arising from non
trivial line bundles and higher order bundles are also present in any global
analysis of duality in quantum field theory, nevertheless, they are usually
missing in the literature.

Higher order bundles \cite{re:MarAlvIsbe}\cite{re:Freed}\cite{re:Nepome}
constitute geometrical objects that generalize the concept of fiber bundles,
from a physical point of view they describe antisymmetric tensor fields with
transitions whose effect is detected in generalized Dirac's quantization
conditions, and are therefore naturally realized in the Chern Simmons terms
in the action for $D=11$ supergravity and consequently relevant for ${\cal {M%
}}$ theory. From the mathematical point of view \cite{re:Brylinski} higher
order bundles -also called Gerbes- are fiber bundles over a manifold, whose
fibers are groupoids.

We begin by reviewing abelian $BF$ theories formulated on trivial bundles.
The action for such systems is given by

\begin{equation}
S=\int{}B\wedge{}dA  \label{eq:S}
\end{equation}

\noindent{}under the assumption that $A$ is a $U(1)$ connection over a
trivial line bundle it follows that $A$ is a globally defined $1$-form over $%
X$ while $B$ is a globally defined $d-2$-form. The field equations following
from (\ref{eq:S}) are:

\begin{equation}
dA=0,\qquad{}dB=0
\end{equation}

\noindent{}that is: $A$ and $B$ are closed $1$ and $d-2$ forms. We define
the gauge transformations for these fields as:

\begin{equation}
B\rightarrow{}d\theta{},\qquad{}A\rightarrow{}d\Lambda  \label{eq:gauge}
\end{equation}

\noindent {}where $\theta $ and $\Lambda $ are a globally defined $d-3$ form
and a $0$-form respectively. The space of solutions of the field equations
coming from the action (\ref{eq:S}) may then be expressed as:

\begin{equation}
H_{dR}^{d-2}(X,\Re{})\otimes{}H_{dR}^{1}(X,\Re{})  \label{eq:alv3}
\end{equation}

\noindent {}where $H_{dR}^{p}(X,\Re {})$ stands for the de Rham Cohomolgy
group of degree $p$. Let us now turn our attention to the quantum theory and
consider the BRST invariant effective action associated to (\ref{eq:S}), to
do so we have to introduce the ghosts, antighosts and Lagrange multipliers
associated to the gauge symmetries (\ref{eq:gauge}) and to the corresponding
BRST gauge fixing procedure. We first notice that the gauge symmetries
correspond to exact forms and that consequently, the associated ghost fields
have no harmonic part. Indeed, since the gauge transformation for the $1$%
-form $A$ is given by

\begin{equation}
A\rightarrow{}A+d\Lambda
\end{equation}

\noindent {}it follows that the corresponding BRST transformation is \cite
{re:kuguh}\cite{re:MODBFV}

\begin{equation}
\hat{\delta}A=dC
\end{equation}

\noindent {}where the zero form $\Lambda $ has been replaced by the ghost
field $C$. Since $d\Lambda $ is an exact form $dC$ must also be exact in
order to preserve the same degrees of freedom, consequently, $dC$ has no
harmonic part. If instead $d\Lambda $ were a closed $1$-form, $dC$ should
also be closed and hence would contain a harmonic part. We also note that
the Lagrange multiplier associated to the gauge fixing introduced by the
modified BFV approach \cite{re:MODBFV} does not have any zero mode freedom
in agreement with Lagrange's multiplier theorem. Consequently the antighost (%
$\bar{C}$) has also no harmonic part contribution. This result can be
generalized to reducible systems (such as $BF$ theories for $d>3$) where one
may argue that the ghosts and antighosts do not have any zero modes
(harmonic parts). In this sense the approach in \cite{re:MODBFV} provides
the construction of the BRST invariant effective action without the
introduction spurious zero modes on the ghost sector. Going back to the
discussion of the quantum theory of the action (\ref{eq:S}) , we find that
the only zero modes are those of the $A$ and $B$ fields which are given by
the cohomology classes (\ref{eq:alv3}), and which give rise to the
nontrivial contributions to the partition function

For the sake of simplicity we consider a straightforward example of the
above discussion, namely, the $BF$ theory in $d=3$ dimensions, the effective
action is given by

\begin{equation}
S_{eff}=\int{}B\wedge{}dA+\lambda_1d^{*}B+\lambda_2\wedge{}D^{*}A+\bar{C}%
_1d^{*}dC_1+ \bar{C}_1d^{*}dC_1  \label{eq:3dabBF}
\end{equation}

\noindent{}by construction there are no zero modes for the ghosts,
antighosts and Lagrange multipliers . Indeed, the potential zero modes for $%
C_{1}$ -for example- arise as solutions of

\begin{equation}
d^{*}dC_1
\end{equation}

\noindent{}but this condition is simply states that $dC_{1}$ is an harmonic $%
1$-form in contradiction with the fact that $dC_{1}$ is exact. The only zero
modes of the effective action (\ref{eq:3dabBF}) come thus from $A$ and $B$.
The zero modes of the $B$ field satisfy

\begin{equation}
dB=0,\qquad{}d^{*}B=0  \label{eq:condsonB}
\end{equation}

\noindent {}and similar conditions for the zero modes of $A$. This set of
conditions imply that the zero modes are two copies of the space of harmonic
$(d-2)=1$-forms. That is, the cohomolgy classes defined by $H_{dR}^{1}(X,\Re
{})\otimes{}H_{dR}^{1}(X,\Re {})$.

The evaluation of the partition function of the $BF$ theory in this case
-over trivial line bundles- was performed in \cite{re:BlauThompson1}\cite
{re:Schwarz} \cite{re:BlauThom2}, the final result is:

\begin{equation}
{\cal {Z}}=\mbox{Vol}({\cal {ZM}})(T(X))^\alpha
\end{equation}

\noindent {}where $\mbox{Vol}({\cal {ZM}})$ denotes the volume of the space
of zero modes, while $T(X)$ is the Ray Singer torsion, the exponent $\alpha $
being given by ($n=dim(X)$):

\begin{equation}
\alpha= \{
\begin{array}{cc}
2-n+1 & n\mbox{ even} \\
-1 & n\mbox{  odd}
\end{array}
\end{equation}

According to the above construction of the effective action, the space of
zero modes ${\cal {ZM}}$ is given by

\begin{equation}
{\cal {ZM}}=H_{dR}^{1}(X,\Re {})\otimes {}H_{dR}^{1}(X,\Re {})
\end{equation}

Concerning the latest point, we take a different point of view than the one
presented in \cite{re:BlauThom2} where the gauge symmetry is extended to
include harmonic gauge parameters, in that case we agree in that the ghosts
sector should include harmonic parts.

We will now turn to our main interest: $BF$ theories defined over non
trivial line bundles. Any connection $1$-form over a nontrivial line bundle
may be decomposed as:

\begin{equation}
A=\hat{A}+a
\end{equation}

\noindent{}where $a$ is a globally defined $1$-form while $\hat{A}$ is a
fixed $1$-form connection of the same topological bundle as $A$. $A$ and $%
\hat{A}$ have the same transitions over $X$, while $a$ has obviously none. $%
B $ will be taken as a globally defined $(d-2)$ form.

The $BF$ action for such a system may be written as:

\begin{equation}
S=i\int_{X}B\wedge{}F(A)=i\int_{X}B\wedge{}F(\hat{A})+i\int_{X}B\wedge{}F(a)
\label{eq:alv11}
\end{equation}

{}Variations of (\ref{eq:alv11}) with respect to $a$, or functional
integration with respect to it yields

\begin{equation}
dB=0  \label{eq:alv12}
\end{equation}
on the other hand, variations of (\ref{eq:alv11}) with respect to $B$ or
equivalently functional integrations on $B$ yields

\begin{equation}
F(A) \equiv dA = 0  \label{eq:alv15}
\end{equation}

\noindent finally, summation over all line bundles gives:

\begin{equation}
\oint_{\Sigma_I}B=2\pi{}n^{I}  \label{eq:alv13}
\end{equation}

\noindent {}where $\{\Sigma _{I}\}$ is a basis of homology of dimension $d-2$%
, while the numbers $n^{I}$ are integers associated to each $\Sigma _{I}.$
To obtain (\ref{eq:alv13}) we recursively integrate by parts using a
triangulation of $X$ and proceeding as follows

\begin{eqnarray}
i\int_{X}B\wedge {}F(\hat{A}) &=&(-1)^{d-2}i\int_{X}d(B\wedge \hat{A}{}%
)=(-1)^{d-2}i\sum_{U_{i}\cap U_{j}}\int_{U_{i}\cap {}U_{j}}B\wedge {}(\hat{A}%
_{i}-\hat{A}_{j})=  \nonumber \\
&=&i(-1)^{d-2}\sum_{U_{i}\cap U_{j}}\int_{U_{i}\cap {}U_{j}}B\wedge d\Lambda
_{ij}=i\sum_{U_{i}\cap U_{j}}\int_{U_{i}\cap {}U_{j}}d(B\Lambda _{ij})=
\nonumber \\
&=&i\sum_{U_{i}\cap U_{j}\cap U_{k}}\int_{U_{i}\cap U_{j}\cap
U_{k}}B(\Lambda _{ij}+\Lambda _{jk}+\Lambda _{ki})=  \nonumber \\
&=&i\sum_{U_{i}\cap U_{j}\cap U_{k}}\int_{U_{i}\cap U_{j}\cap U_{k}}2\pi
nB=2\pi m^{I}\sum_{I}i\int_{_{\Sigma ^{I}}}B
\end{eqnarray}

\noindent {}summation over all $m^{I}$ yields then (\ref{eq:alv13}). In (\ref
{eq:alv11}) one may identify the last term of the split action, namely: $%
i\int_{X}B\wedge {}F(a)$ as the $BF$ action of a global $1$-form $a$ or
equivalently of a $1$-form connection over a trivial line bundle. The
contribution of all non-trivial $U(1)$ line bundles arises from the $%
i\int_{X}B\wedge {}F(\hat{A})$ term of the action.

\noindent {}Let us now analyze the space of solutions of the field equations
(\ref{eq:alv12}) ,(\ref{eq:alv15}), and (\ref{eq:alv13}). We first recall
that $F(A)$ being the curvature of a connection $1$-form $A$ over a line
bundle also has integral periods (Dirac quantization conditions) over any
basis of homology of dimension $2$, but because of (\ref{eq:alv15}) the
integers are all zero. Condition (\ref{eq:alv15}) implies that $A$ is a flat
connection $1$-form, given $A_{flat}$ one such connection then for any
closed $1$-form $\omega $ over $X$

\begin{equation}
A_{flat}+\omega  \label{eq:alv16}
\end{equation}

\noindent{}is also a flat connection on the same line bundle, moreover these
are all the flat connections over the line bundle. The problem then reduces
to find all the line bundles which admit a flat connection. There exists
generically non-trivial line bundles with flat connections over it. They are
line bundles with constant transitions and are classified by the $\check{C}%
ech$ cohomology group with values on the constant sheaf $\Re{}/Z$:

\begin{equation}
\check{H^{1}}(X,\Re{}/Z)  \label{eq:alv17}
\end{equation}

For any such line bundle the flat connections are in one to one
correspondence to the de Rahm cohomology group

\begin{equation}
H_{dR}^{1}(X,\Re {}).  \label{eq:alv18}
\end{equation}

\noindent{}According to this, the total space of flat connections is
determined by the product

\begin{equation}
\check{H^{1}}(X,\Re /Z)\otimes H_{dR}^{1}(X,\Re )  \label{eq:alv19}
\end{equation}

Let us now analyze the space of solutions to (\ref{eq:alv12}) and (\ref
{eq:alv13}), taking into account the gauge symmetry of the $B$ field

\begin{equation}
B\rightarrow B+d\theta  \label{eq:alv20}
\end{equation}

\noindent {}where $d\theta $ is an exact $(d-2)$ form. Conditions $dB=0$ and
$\oint_{\Sigma _{I}}B=2\pi {}n^{I}$ ensure the existence of both: a higher
order $U(1)$ bundle, and a $(d-3)$-form $b$ with non-trivial transitions
over $X$ \cite{re:MarAlvIsbe} , such that $B=db$. The former result is a
generalization of Weil's theorem stating that given a $2$-form $F$
satisfying Dirac's quantization conditions there exists a line bundle and a
connection $A$ such that $F$ is the curvature of $A$ \cite{re:Brylinski}

In a particular case, $B$ being a $3$-form we have, on an open covering $%
U_{l},l\in L$, of $X$ that

\begin{eqnarray}
&B=db_{i}\mbox{ on }U_{i}\mbox{,}\qquad {}b_{i}-b_{j}=d\eta _{ij}\mbox{ on }%
U_{i}\wedge U_{j}\neq \emptyset &  \nonumber \\
&\eta _{ij}+\eta _{jk}+\eta _{ki}=d\Lambda _{ijk}\mbox{ on }U_{i}\wedge
U_{j}\wedge U_{k}\neq \emptyset &  \label{eq:alv21} \\
&\displaystyle{\sum_{ijkl}}\Lambda _{ijkl}=2\pi {}n&  \nonumber
\end{eqnarray}

\noindent {}the cocycle condition being satisfied on any intersection of
four open sets. Conversely, given conditions (\ref{eq:alv21}), then $B$
satisfies (\ref{eq:alv12}) and (\ref{eq:alv13}).

The gauge equivalent triplets $(b,\eta {},\Lambda {})$ are defined by

\begin{eqnarray}
&b_{i}\rightarrow b_{i}+d\eta _{i}\qquad \mbox{on }U_{i}&  \nonumber \\
&\eta _{ij}\rightarrow \eta _{ij}+\eta _{i}-\eta _{j}+d\Lambda _{ij}\qquad
\mbox{on
}U_{i}\cap U_{j}&  \label{eq:alv22} \\
&\Lambda _{ijk}\rightarrow \Lambda _{ijk}+\Lambda _{ij}+\Lambda
_{jk}+\Lambda _{ki}\qquad \mbox{on }U_{i}\cap U_{j}\cap U_{k}&  \nonumber
\end{eqnarray}

\noindent {}It is important to realize that given $B$ satisfying (\ref
{eq:alv12}) and (\ref{eq:alv13}) the triplet $(b,\eta {},\Lambda {})$ in (%
\ref{eq:alv21}) may not be unique in general, indeed, there may exist
constant transition $\widetilde{\Lambda }_{ijk}$ on $U_{i}\cap U_{j}\cap
U_{k}\cap U_{l}\neq \emptyset $ satisfying

\begin{equation}
\sum_{ijkl}\widetilde{\Lambda }_{ijk}=0\hspace{10pt}\mbox{on}\hspace{10pt}%
U_{i}\cap U_{j}\cap U_{k}\cap U_{l}\neq \emptyset
\end{equation}

\noindent {}which may be added to any particular triplet $(b,\eta {},\Lambda
{})$ satisfying (\ref{eq:alv21}) giving rise to a new triplet satisfying the
same conditions. The space of all the triplets $(0,0,\widetilde{\Lambda }%
_{ijk})$ with constant transitions $\widetilde{\Lambda }_{ijk}$are
classified by the $\check{C}ech$ cohomology group $\check{H^{2}}(x,\Re /Z)$
with values on the constant sheaf $\Re /Z$. This is the only degeneracy on
the triplets satisfying (\ref{eq:alv21}) for a given $B$, a closed $3$-form
with integral periods. Modulo this degeneracy two triplets $(b_{1},\eta
_{1},\Lambda _{1})$ and $(b_{2},\eta _{2},\Lambda _{2})$ satisfying the
cocycle condition with the same set of integers $n$ satisfy

\begin{equation}
(b_{2},\eta _{2},\Lambda _{2})\sim (b_{1}+\theta ,\eta _{1},\Lambda _{1})
\end{equation}

\noindent {}where $\sim $ denotes gauge equivalence (\ref{eq:alv22}), and $%
\theta $ is a globally defined $2$-form over $X$. Since this is just the
gauge symmetry (\ref{eq:alv20}) of the $BF$ action and consequently of its
field equations (\ref{eq:alv12}) and (\ref{eq:alv13}), we conclude that the
set of integers $n^{I}$ associated to a basis of integral homology of
dimension $3$ determine the space of solutions of (\ref{eq:alv12}) and (\ref
{eq:alv13}). They classify all the higher order bundles of degree $3$, that
is the $\check{C}ech$ cohomology group

\begin{equation}
\check{H^{3}}(X,Z)  \label{eq:alv25}
\end{equation}

\noindent modulo $\check{H^{2}}(X,\Re {}/Z)$ the space of higher order
bundles of degree $3$ with constant transitions, i.e. two elements of the
same equivalence class differ by constant transitions $\widetilde{\Lambda }.$

\noindent Finally, from (\ref{eq:alv11}) we may determine the partition
function of the $BF$ theory on non-trivial line bundles. It has the form $%
\mbox{Vol}({\cal {ZM}})(T(X))^{\alpha }$ already found in \cite
{re:BlauThompson1}\cite{re:Schwarz}\cite{re:BlauThom2} but now the zero mode
space is determined by

\begin{equation}
(\check{H^{1}}(X,\Re /Z)\otimes {}H_{dR}^{1}(X,\Re {}))\otimes {}(\check{%
H^{3}}(X,Z)/\check{H^{2}}(X,\Re {}Z)).
\end{equation}

In the general case the degree of the $\check{C}ech$ cohomology groups are $%
d-2$ and $d-3$ respectively and consequently the last formula generalizes to

\begin{equation}
(\check{H^{1}}(X,\Re /Z)\otimes {}H_{dR}^{1}(X,\Re {}))\otimes {}(\check{%
H^{d-2}}(X,Z)/\check{H^{d-3}}(X,\Re {}Z)).
\end{equation}

\noindent showing that $BF$ theories provide, may be the most elementary
realization of higher order bundles in field theory.


\begin{thebibliography}{99}
\bibitem{re:Horowitz}  G. Horowitz, Commun. Math. Phys. 125 (1984) 417.

\bibitem{re:BlauThompson1}  M. Blau and G. Thompson, Ann. Phys. 205 (1991)
130.

\bibitem{re:Schwarz}  A. Schwarz, Lett. Math. Phys. 2 (1978) 247.

\bibitem{re:BlauThom2}  D. Birmingham, M. Blau, M. Rakowski and G. Thompson,
Phys. Rep. 209 (1991) 129.

\bibitem{re:MarAlv1}  M. I. Caicedo and A. Restuccia, Phys. Lett. B307
(1993) 77.

\bibitem{re:MarAlv2}  M. I. Caicedo, R. Gianvittorio, A. Restuccia and J.
Stephany Phys. Lett. B354 (1995) 292.

\bibitem{re:PariasR}  P. J. Arias and A. Restuccia Phys.Lett. B347 (1995)
241.

\bibitem{re:Stephany}  J. Stephany, Phys. Lett. B390 (1997), 128.

\bibitem{re:MarAlvIsbe}  M. I. Caicedo, I. Martin and A. Restuccia, {\it %
Duality on Higher Order Bundles}, hep-th/9701010. {\it On the Geometry of
Antisymmetric Fields}, hep-th/9711122.

\bibitem{re:Freed}  D. S. Freed, talk given at the XIX International
Colloquium on Group Theoretical Methods in Physics, Salamanca 1992.

\bibitem{re:Nepome}  P. Freund and R. Nepomechie, Nuc. Phys. {\bf B199}
(1982) 482, R. Nepomechie, Phys. Rev. {\bf D31}. (1985), 1921.

\bibitem{re:Brylinski}  Jean-Luc Brylinski, {\it Loop Spaces, Characterisic
Classes and Geometric Quantization}, Birkh\"{a}user, 1992.

\bibitem{re:kuguh}  T. Kugo and S. Uehara, Nuc. Phys. {\bf B197} (1982) 378.

\bibitem{re:MODBFV}  M. I. Caicedo and A. Restuccia, Class.Quant.Grav.10
(1993) 833.
\end{thebibliography}
\end{document}